\documentclass[prd,twocolumn,showpacs,preprintnumbers,amsmath,amssymb]{revtex4}

\usepackage{graphicx}
\usepackage{dcolumn}
\usepackage{bm}

\newcommand{\order}{{\cal O}}

\newcommand{\be}{\begin{equation}}  
\newcommand{\ee}{\end{equation}}  
\newcommand{\bear}{\begin{eqnarray}}  
\newcommand{\eear}{\end{eqnarray}}  
\newcommand{\ba}{\begin{array}}  
\newcommand{\ea}{\end{array}}

\newskip\humongous \humongous=0pt plus 1000pt minus 1000pt

\newif\ifdtup

\def\oldreffmt#1{\rlap{[#1]} \hbox to 2\parindent{}}

\def\figfmt#1{\rlap{Figure {#1}} \hbox to 1in{}}  
  
\def\beq{\begin{equation}}  
\def\eeq{\end{equation}}  
\def\bea{\begin{eqnarray}}  
\def\eea{\end{eqnarray}}

\def\bq{\begin{quote}}  
\def\eq{\end{quote}}

\relax  

\newdimen\tdim  
\tdim=\unitlength  
\def\bar{\overline}

\begin{document}  

\preprint{EFI Preprint 10-3}
\title{
On the single photon background to $\nu_e$ appearance 
at MiniBooNE
}

\author{
Richard J. Hill
}

\email{
richardhill@uchicago.edu
}

\affiliation{\vspace{0.2in}
Enrico Fermi Institute and Department of Physics \\
The University of Chicago, Chicago, Illinois, 60637, USA
}

\date{\today}

\begin{abstract} 
Neglected single photon processes are fit to an excess of
electron-like events observed in a predominantly $\nu_\mu$ 
beam at MiniBooNE.  Predictions are given for analogous events in 
antineutrino mode.  
\end{abstract}

\pacs{
12.38.Qk, 
12.39.Fe, 
13.15.+g, 
}
\maketitle

\section{Introduction}
The MiniBooNE experiment was designed to test 
the indication of a $\bar{\nu}_\mu \to \bar{\nu}_e$ oscillation signal at 
LSND~\cite{Athanassopoulos:1996jb,Athanassopoulos:1997pv}.  
MiniBooNE data for $\nu_e$ appearance in a $\nu_\mu$ 
beam~\cite{AguilarArevalo:2007it}, when restricted to  
the range of $475-1250\,{\rm MeV}$ reconstructed neutrino 
energy, refute a simple two-neutrino oscillation interpretation for the LSND signal.  
However, the results indicate an excess of signal-like events at low energy, 
which has persisted at the $3\sigma$ level after various refinements to the analysis~\cite{AguilarArevalo:2008rc}.  
First results from MiniBooNE for $\bar{\nu}_e$ appearance 
in a $\bar{\nu}_\mu$ beam do not show a significant excess~\cite{AguilarArevalo:2009xn}, 
but are inconclusive with respect to the LSND signal. 

Because electromagnetic showers instigated by electrons and photons
are not distinguishable at MiniBooNE, neutral current events producing
single photons are an irreducible background to the charged 
current $\nu_e n \to e^- p$ signal.   
This note presents flux-averaged cross sections 
for the dominant sources of single-photon backgrounds, some of which
were not incorporated in the MiniBooNE analysis.  
These standard model processes must be well-understood and accounted for 
before appealing to more exotic interpretations of the electron-like signal.

\section{Single photon processes}

The $\sim 1\,{\rm GeV}$ energy range is unfortunately not well 
suited to precise analytic results, since there is
no obvious small expansion 
parameter for this regime of QCD~\footnote{The discussion can be formalized by 
appealing to the large $N_c$ limit.}. 
At low energy, contributions to the process of
interest can be tabulated in the rigorous language 
of a chiral lagrangian expansion.   
Extrapolation to moderate energy can then be performed by 
explicitly including the lowest-lying resonances in each channel,
and adopting phenomenological form factors to mimic the effects of higher resonances.  
This methodology was discussed in detail in \cite{Hill:2009ek}. 
This study was motivated by the search for
low-energy remnants of the baryon current anomaly, such 
as the coherent coupling of weak and electromagnetic currents to 
baryon density~\cite{HHH2}\footnote{
In terms of the baryon chiral lagrangian, the operator induced by 
$t$-channel $\omega$ exchange (considered in \cite{HHH2}) also receives
contributions from $s-$ and $u$-channel $\Delta$.  The amplitudes constructively 
interfere at energies $E\ll m_\omega, m_\Delta-m_N$, with $\Delta$ appearing to 
give the larger contribution.   
Further discussion, including extrapolation to larger 
energy and kinematic distributions for different photon production mechanisms, can 
be found in \cite{Hill:2009ek}.  
}.

In terms of the chiral lagrangian expansion, 
the production of single photons in neutrino scattering on 
nucleons begins at order $1/M$~\footnote{
$M\sim m_N \sim 4\pi f_\pi \sim 1\,{\rm GeV}$.
}.
The $1/M$ contributions represent offshell intermediate 
nucleon states or Compton-like scattering, including bremsstrahlung corrections to elastic scattering. 
At the next order there appears a term that derives from 
$s$-channel $\Delta(1232)$ production and $t$-channel $\omega(780)$ exchange.   
Exchange of $\pi^0$ in the $t$-channel is naively of similar size but is suppressed by an
amplitude factor $1-4\sin^2\theta_W \approx 0.08$~\cite{Jenkins:2009uq}.   Similarly, 
exchange of isovector $\rho(770)$ is suppressed relative to isoscalar 
$\omega(780)$ by a factor $\sim (1/3)^2$ from quark counting rules. 
The $\Delta$ and $\omega$ contributions are spin-independent interactions
at low energy, and can also give rise to coherent scattering on compound 
nuclei such as $^{12}C$ in the MiniBooNE detector.  
The coherent contribution from bremsstrahlung emission on the nucleus is 
numerically small.  

In what follows, the coherent bremsstrahlung process, and the incoherent $\pi^0$ and $\rho^0$ 
processes, are neglected. 
The remaining contributions are computed using the parameter values and form factor models from \cite{Hill:2009ek}
and the published MiniBooNE fluxes in both neutrino and antineutrino 
modes~\cite{AguilarArevalo:2008yp}.  
Only the incoherent $\Delta$ contribution was studied in the MiniBooNE 
analysis, with normalization constrained by comparison to observed $\pi^0$ production and a model of 
final state interactions.   
The result of this procedure is compared to a direct calculation of the incoherent $\Delta$ contribution.  
Nuclear effects and other uncertainties are briefly discussed. 

\section{MiniBooNE neutrino cross sections} 

\begin{figure}[h]
\begin{center}
\includegraphics[width=20pc, height=15pc]{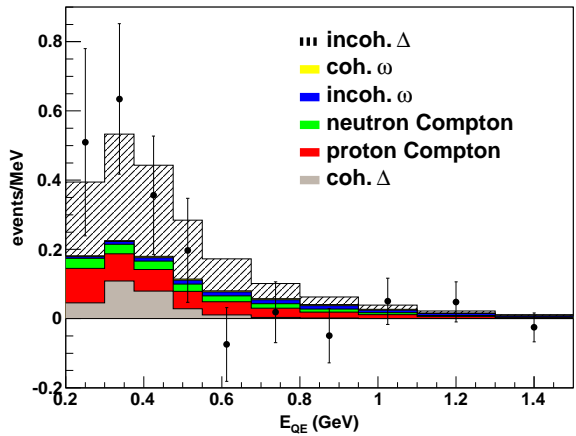}
\caption{\label{fig:nu_e_qe}
Single-photon events at MiniBooNE for $6.46\times 10^{20}$ protons on target in neutrino mode.  
A $25\%$ efficiency is assumed.   The hatched line represents the difference
between the direct calculation and MiniBooNE $\pi^0$-constrained incoherent 
$\Delta\to N\gamma$ background.   Data points correspond to the excess 
events reported in \cite{AguilarArevalo:2008rc}, Fig.~2. 
}
\end{center}
\end{figure}

\begin{figure}[h]
\begin{center}
\includegraphics[width=20pc, height=15pc]{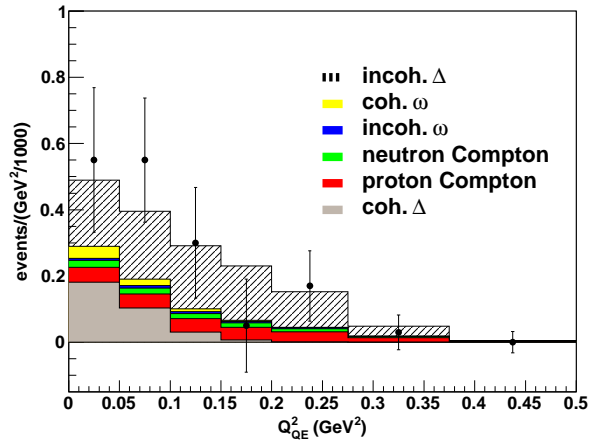}
\includegraphics[width=20pc, height=15pc]{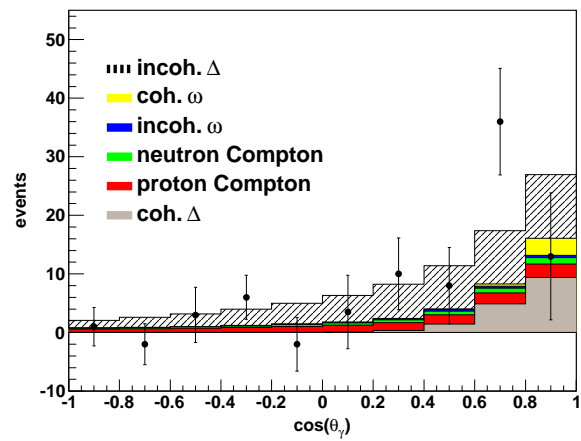}
\caption{\label{fig:nu_dist}
Distributions in $Q^2_{QE}$ and $\cos\theta$ for the events displayed
in Figure~\ref{fig:nu_e_qe} for $E_{QE}=300-475\,{\rm MeV}$.  Data points 
correspond to Figs.~4 and 5 of \cite{AguilarArevalo:2008rc}.
}
\end{center}
\end{figure}

Figure~\ref{fig:nu_e_qe} displays flux-integrated
cross sections, presented as events per MeV of reconstructed neutrino energy~\footnote{
The reconstructed neutrino energy 
assumes charged current quasielastic scattering on a stationary nucleon 
to an electron final state.   Neglecting the electron mass, we have  
$E_{\rm QE} \equiv E_{\rm vis}/[1- (E_{\rm vis}/m_N)(1-\cos\theta)]$. 
}.   The normalization corresponds to a 
detector mass of $800\times 10^6 g$, and $6.46\times 10^{20}$ protons on target 
for the updated analysis of $\nu_e$ charged current quasi-elastic (CCQE) events in a primarily $\nu_\mu$ 
beam~\cite{AguilarArevalo:2008rc}.  
A cut $E_{\gamma}\ge 140\,{\rm MeV}$ is placed on the photon energy, in accordance
with the experimental selection.  

To compare to the MiniBooNE data  in the absence of a dedicated efficiency analysis,  
the number of events has been multiplied by an efficiency factor of $25\%$ and detector resolution/smearing 
effects have been neglected.  
For comparison, the original MiniBooNE analysis quotes an efficiency of $30.6\pm 1.4\%$ for reconstructing  
signal-like $\nu_e$ CCQE events~\cite{AguilarArevalo:2007it}.  
As can be seen from Table~\ref{table:nu}, after selection cuts the efficiency for 
events with similar signatures, $\nu_\mu e^- \to \nu_\mu e^-$ and $\nu_e n \to e^- p$, 
fall in the range $20-30\%$~\footnote{
For the CCQE events, the small $\bar{\nu}_e$ contamination in the beam is neglected and the 
cross section is calculated in a standard relativistic Fermi gas model of Smith and Moniz~\cite{Smith:1972xh},
with model parameters $\epsilon_b=27\,{\rm MeV}$ and $k_F=225\,{\rm MeV}$. 
}. 
It can also be seen from this table that the 
direct estimate of the number of single photon events mediated by $\Delta(1232)$ is larger 
than the $\pi^0$-constrained background estimate of MiniBooNE by a factor $\approx 2$~\footnote{
Using the energy-dependent Breit-Wigner width gives a lower total cross section and 
pulls events to lower energy: with the same form factors the number of events would be 217, 363 and 
175 in the $200-300$, $300-475$ and $475-1250\,{\rm MeV}$ bins, respectively.  For definiteness, we 
focus on the ``default'' cross sections from \cite{Hill:2009ek}.  
}.  
The effects of a larger incoherent $\Delta\to N\gamma$ background are illustrated by 
the hatched area in Fig.~\ref{fig:nu_e_qe}, computed by adding $0.5$ times the direct estimate
(i.e., effectively doubling the MiniBooNE background).  
Under the assumption of a constant $25\%$ efficiency, the fit of these additional single-photon events
to the MiniBooNE excess yields $\chi^2=10.3$ for 10 d.o.f.   
Theoretical errors are discussed at the end of this note and have not been included in the fit.  
Assuming a lower $20\%$ efficiency
and taking the difference between the estimates of $\Delta\to N\gamma$ events from the table, 
the remaining excess would be $15\pm 26$, $23\pm 25$ and $-47\pm 36$ in the 
$200-300$, $300-475$ and $475-1250\,{\rm MeV}$ bins, respectively.   
If no  additional incoherent $\Delta\to N\gamma$ 
events are included, these numbers become $29\pm 26$, $55\pm 25$ and $-9\pm 36$.  

\begin{table}
\caption{\label{table:nu}
Single photon and other backgrounds for MiniBooNE $\nu$-mode in ranges of $E_{\rm QE}$. 
Ranges in square brackets are the result of applying a $20-30\%$ efficiency 
correction.
}
\begin{ruledtabular}
\begin{tabular}{cccc}
process & 200-300 & 300-475 & 475-1250 \\
\hline
1$\gamma$, non-$\Delta$ & $85[17-26]$ & $151[30,45]$ & $159[32,48]$ \\ 
$\Delta\to N\gamma$ & $170 [34-51]$ & $394[79-118]$ & $285[57-86]$ \\
$\nu_\mu e \to \nu_\mu e$ & $14 [2.7-4.1]$ & $20[4.0-5.9]$ & $40[7.9-12]$ \\
$\nu_e n \to e p$ & $100[20-30]$ & $303[61-91]$ & $1392[278-418]$ \\
\hline
MB excess & $45.2\pm 26.0$ & $83.7 \pm 24.5$ & $22.1\pm 35.7$ \\
MB $\Delta\to N\gamma$ & 19.5 & 47.5 & 19.4 \\ 
MB $\nu_\mu e \to \nu_\mu e$ & 6.1 & 4.3 & 6.4 \\
MB $\nu_e n \to e p$ & 19 & 62 & 249 
\end{tabular}
\end{ruledtabular}
\end{table}

The most significant excess in the updated MiniBooNE 
analysis occurred in the $E_{\rm QE} = 300-475\,{\rm MeV}$ bin.
The distributions in reconstructed $Q^2$~\footnote{
The reconstructed $Q^2$ variable is defined, neglected electron mass, as 
$Q^2_{\rm QE} \equiv 2E_{\rm vis}^2 (1-\cos\theta)/[1-(E_{\rm vis}/m_N)(1-\cos\theta)]$.
},
and cosine of the angle, $\cos\theta$, of the electromagnetic shower with respect to the
beam direction, are displayed for this energy range in Figure~\ref{fig:nu_dist}.
The normalization assumes an energy- and angle-independent 
efficiency of $25\,\%$, and includes $0.5$ times the incoherent $\Delta\to N\gamma$ background 
as in Figure~\ref{fig:nu_e_qe}. 
A $\chi^2$ fit yields $10.9/ {\rm 10\, d.o.f.}$ for $\cos\theta$  
and $2.6/{\rm 7 \, d.o.f.}$ for  $Q^2_{QE}$.  

Note that in the accounting method here, it does not matter whether the MiniBooNE $\Delta\to N\gamma$
background estimate represents just the incoherent, or the sum of incoherent plus coherent processes.
In the latter case, the difference between the $\pi^0$-constrained background and the direct estimates
given here would be larger; the ``$\Delta$'' and ``coherent $\Delta$'' regions in the figures would 
contribute different amounts but with the same total.

From the estimates presented here, it may be difficult
to extract the coherent component from other backgrounds.  Doing so would represent the first 
signal for coherent single photon production by the weak neutral current above the nuclear scale~\footnote{
Nuclear excitation yields coherent photon production at the $\sim {\rm MeV}$ energy scale, 
e.g.~\cite{Armbruster:1998gk}. 
While some hints of single photon events were discussed in previous high-energy experiments, these
were not definitively isolated as being due to single photons as opposed e.g. to 
$\pi^0$ production with a missed photon~\cite{Faissner:1978qu,Isiksal:1982sc}.
}.

\section{MiniBooNE antineutrino cross sections}

\begin{figure}
\begin{center}
\includegraphics[width=20pc, height=15pc]{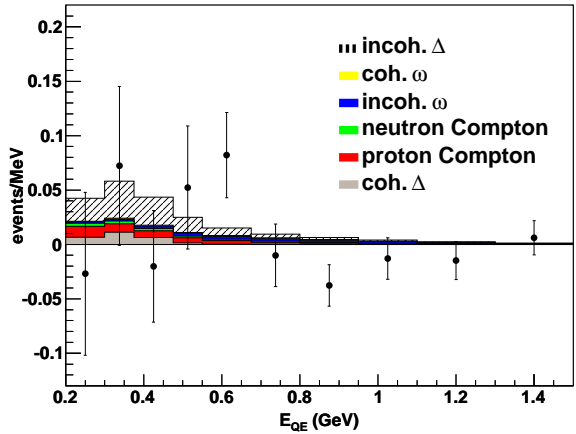}
\caption{\label{fig:nubar_excess_E_QE}
Comparison of single photon events to MiniBooNE data with other backgrounds
subtracted in antineutrino mode.  
}
\end{center}
\end{figure}

The above procedure may be repeated for antineutrinos.  
Figure~\ref{fig:nubar_excess_E_QE} displays flux-integrated cross sections
normalized according to $3.39\times 10^{20}$ protons on target   from
the search for $\bar{\nu}_e$ CCQE events in a 
primarily $\bar{\nu}_\mu$ beam~\cite{AguilarArevalo:2009xn}. 
A cut $E_{\gamma}\ge 140\,{\rm MeV}$ is applied, and a  
$25\%$ efficiency has been assumed, in accordance with a 
comparison to MiniBooNE backgrounds in Table~\ref{table:nubar}~\footnote{
For CCQE events, the estimate includes $4/3$ of the $\bar{\nu}_e$ cross section on $^{12}C$
to account for hydrogen (neglecting the difference between free and bound protons), as well
as the $\nu_e$ cross section on $^{12}C$.  Cross sections are computed as in the $\nu_e$ case, 
using a relativistic Fermi gas model. 
}.  
Again, the direct estimate of $\Delta\to N\gamma$ events is $\approx 2$ times larger
than the MiniBooNE estimate; the difference is illustrated in the figure by including
$0.5$ times the direct estimate for these events.    The resulting fit for the $E_{QE}$ 
distribution yields $\chi^2=13.3$ for 10 d.o.f.    
Assuming a $20\%$ efficiency
and taking the difference between the estimates of $\Delta\to N\gamma$ events from the table, 
the excess becomes $-11.5\pm 11.7$ and $-2.8\pm 10.0$ in the $200-475$ and $475-1250\,{\rm MeV}$ 
bins, respectively.   If no additional incoherent $\Delta\to N\gamma$ events are included, these numbers
become $-6.1\pm 11.7$ and $-0.2\pm 10.0$.  

\begin{table}
\caption{\label{table:nubar}
Single photon and other backgrounds for MiniBooNE $\bar{\nu}$-mode in ranges of $E_{\rm QE}$. 
Ranges in square brackets are the result of applying a $20-30\%$ efficiency 
correction.
}
\begin{ruledtabular}
\begin{tabular}{cccc}
process & 200-475 & 475-1250 \\
\hline
1$\gamma$, non-$\Delta$ & 28[5.6-8.4] & 17[3.4-5.2] \\
$\Delta\to N\gamma$ & 58[12-17] & 23[4.6-6.9] \\ 
$\bar{\nu}_e/\nu_e\,{\rm CCQE}$ & 81[16-24] & 261[52-78] \\
\hline
MB excess & $-0.5\pm 11.7$ & $3.2\pm 10.0$ \\ 
MB $\Delta\to N\gamma$ & 6.6 & 2.0 \\
MB $\bar{\nu}_e/\nu_e\,{\rm CCQE}$ & 18 & 43 
\end{tabular}
\end{ruledtabular}
\end{table}

\section{Nuclear effects and other uncertainties}

\begin{figure}
\begin{center}
\includegraphics[width=20pc, height=15pc]{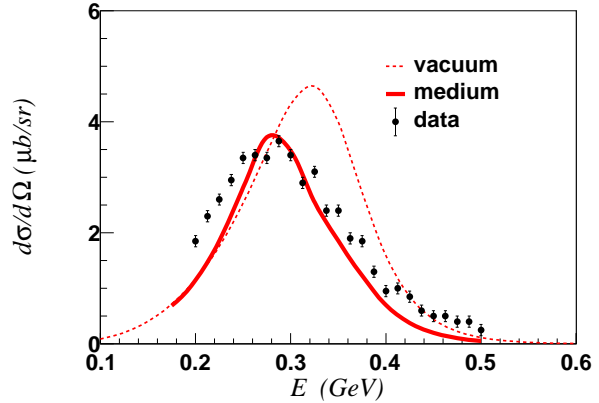}
\caption{\label{fig:compton_energy}
Coherent photon scattering on $^{12}C$ mediated by $\Delta$ resonance 
at $\theta=40^\circ$ ($\cos\theta=0.766$).    Dashed line is model with 
energy-independent width $\Gamma_\Delta=120\,{\rm MeV}$.  Solid
line is for energy-dependent width and in-medium effects as described in
text.   Data is from \cite{Wissmann:1994tj}. 
}
\end{center}
\end{figure}

The absence of an obvious small expansion parameter complicates error 
estimation for $E\sim{\rm GeV}$ cross sections involving hadronic matrix elements.  
At the single nucleon level, the $1/N_c$ expansion of 
QCD motivates the assignment of $\order(1/N_c)\sim 30\%$ uncertainty to 
tree level amplitudes when all of the relevant resonances are included.  
Nuclear corrections induce additional uncertainty.  
Here we consider some of 
these, focusing on 
in-medium effects on the coherent cross section, and
Fermi motion effects on the incoherent cross section. 

The largest component of the coherent amplitude at low energy is due to 
the $\Delta(1232)$ resonance 
in the $s$ channel~\footnote{This process may be viewed as 
a ``reverse coherent GZK'' effect~\cite{Greisen:1966jv}.}.  
To gauge whether the nuclear modeling for the coherent process 
mediated by $\Delta$ excitation gives a reasonable 
approximation to the true cross section, 
it is useful to compare to the analog process of coherent photon scattering on 
the same nucleus.  Data from \cite{Wissmann:1994tj}
at a fixed angle $\theta=40^\circ$ is shown in Fig.~\ref{fig:compton_energy}. 
The dashed line in the figure shows the result of the ``default'' model from \cite{Hill:2009ek}
(with photon in place of the vector-coupled $Z^0$), 
using energy-independent width $\Gamma_\Delta=120\,{\rm MeV}$.    
For comparison, the result of including in-medium modifications to 
$\Delta$ propagation and using energy-dependent width
is displayed as the solid line.  Here a simple model for these modifications is taken from Drechsel 
et.al.~\cite{Drechsel:1999vh}~\footnote{
The modifications are implemented using Eqs.(25) and (26a) and Fig.5 of \cite{Drechsel:1999vh}.
The complex phase $\phi$ discussed there has been neglected. 
}.

As the figure illustrates, the data is in better agreement with the model 
incorporating in-medium effects, where the cross section is somewhat reduced, and
the peak shifted to smaller energy.  
The fit to the data can be improved by using a slightly larger vacuum width 
(e.g. $\Gamma_\Delta\sim 130\,{\rm MeV}$) and including a small nonresonant background. 
However, these modifications are beyond the accuracy of other approximations 
such as the simplified nuclear form factor and nonrelativistic reduction of the amplitude. 
The main point to be illustrated is that the simple model represented by the dashed 
line is not a gross misrepresentation of the data.   

The incoherent single-photon cross sections have been calculated neglecting nuclear effects. 
From a comparison of the CCQE cross sections computed assuming free nucleons, and those using 
a relativistic Fermi gas model, the nuclear binding, Fermi motion of the initial-state nucleons, 
and Pauli blocking of the final state nucleons,
are seen to affect the cross sections on the $\sim 10-15\%$ level for the relevant energies.   
Similarly, in-medium modification of parameters such as $m_\Delta$, $\Gamma_\Delta$ should
affect the cross sections at a $\sim 10-20\%$ level, as seen in the above example for coherent
Compton scattering, or in related analyses of single pion production through the $\Delta$ 
resonance~\footnote{See e.g. \cite{Leitner:2008wx}.  In this case, the final-state interactions of the pion 
are intertwined with in-medium modifications to the primary vertex.}.
The effects of higher baryon resonances are small ($< 10\%$) relative to $\Delta(1232)$,
based on the production cross sections  below $2\,{\rm GeV}$ neutrino energy~\cite{Lalakulich:2006sw}.   
Interference effects between different photon production mechanisms ($\omega$, $\Delta$, Compton) 
have been neglected.  
A detailed study is beyond the scope of this work.  The distinct kinematic distributions~\cite{Hill:2009ek} 
suggest that interference effects will not drastically alter the total cross sections at the
energies accessible to MiniBooNE.  Photons produced by rescattering of pions in the nucleus have been neglected.

The range of $30-50\%$ is a subjective estimate of the total single-photon cross section uncertainty. 
A dedicated efficiency analysis would constrain the overall normalization error. 
Examination of processes such as $\nu_\mu n \to \mu^- p \gamma$ could be used to test other
sources of uncertainty~\footnote{A sample of muon events with photons was 
isolated in \cite{:2007hy}.  J. Conrad, private communication.}.  
This process avoids complications from final-state pion interactions, but 
involves different linear combinations of the hadronic matrix elements than contribute 
to the neutral current process.

\section{Summary}

Neglected single photon events give a significant contribution to the MiniBooNE 
low-energy excess.  
Fits to the data also favor an enhanced resonant 
$\Delta\to N\gamma$ contribution (either incoherent or coherent) relative to estimates 
based on $\pi^0$ production.   
A similar enhancement is predicted by a phenomenological model calculation, and is consistent 
with the absence of a significant excess in the MiniBooNE antineutrino results. 
An enhanced coupling of the neutral weak current and electromagnetic current 
to baryons may have interesting astrophysical implications~\cite{HHH2}. 

\section{Acknowledgements}

I have benefited from discussions with M.~Shaevitz,  J.~Conrad, 
G.~Garvey, J.~Harvey, C.~Hill, G.~Paz and G.~Zeller. 
Work supported by NSF Grant 0855039.

\end{document}